\documentclass[12pt, letterpaper, titlepage]{article}
\pdfoutput=1
\usepackage{amsmath}
\usepackage{booktabs}
\usepackage{amsthm}
\usepackage{graphicx}
\usepackage[margin=1in]{geometry}
\usepackage {hyperref}
\hypersetup {colorlinks = true, linkcolor = blue, citecolor=blue, urlcolor = blue}
\usepackage[]{natbib}
\usepackage{enumitem}
\usepackage{setspace}

\title{The Effects of the NBA COVID Bubble on the NBA Playoffs: A Case Study for Home-Court Advantage}

\author{Michael Price and Jun Yan\\
\href{mailto:michael.price@uconn.edu}{\nolinkurl{michael.price@uconn.edu}}\\
Department of Statistics, University of Connecticut}
\date{}

\begin{document}
\maketitle

\doublespace

\begin{abstract}
The 2020 NBA playoffs were played inside of a bubble in Disney World because of
the COVID-19 pandemic. This meant that there were no fans in attendance,
games played on neutral courts and no traveling for teams, which in theory
removes home-court advantage from the games. This setting has attracted much
discussion as analysts and fans debated the possible effects it may have on the
outcome of games. Home-court advantage has historically played an influential
role in NBA playoff series outcomes. The 2020 playoff provided a unique
opportunity to study the effects of the bubble and home-court
advantage by comparing the 2020 season with the seasons in the past.
While many factors contribute to the outcome of games, points scored
is the deciding factor of who wins games, so scoring is the primary focus
of this study. The specific measures of interest are team scoring totals and
team shooting percentage on two-pointers, three-pointers, and free throws.
Comparing these measures for home teams and away teams in 2020 vs.~2017-2019
shows that the 2020 playoffs favored away teams more than usual, particularly
with two point shooting and total scoring.
\end{abstract}

\hypertarget{sec:intro}{%
\section{Introduction}\label{sec:intro}}

Home-court advantage is often discussed in sports circles as a contributing
factor to the outcome of games. It is well-known that the home team typically
benefits from some competitive edge from playing at their home-court, resulting
in a better chance of winning. Thus, the NBA playing the 2020 playoffs in a
bubble due to the COVID-19 pandemic
brought a great deal of concern for fans, teams, journalists, and others.
For example, \citet{Aschburner} discusses about anticipated effects, sharing
concerns from former players, coaches and other experts about
the potential effects of removing home-court advantage. Aschburner notes that the
NBA did make attempts to recreate the effects by putting the ``home'' team logo on
the court and allowing the ``home'' team to play crowd noise and music, but most
people doubted these small attempts would recreate a true playoff atmosphere.
During the 2020 NBA playoffs, home teams only won about 48.2\% of the games. This
is lower than normal, which Aschburner claims usually floats around 60\%. This
shift in the home team winning percentage surely indicates the opportunity for
thorough investigation.

So, what happened? Did the home teams fail to perform up to normal standards
without the help of home-court advantage? Were away teams able to rise to
the occasion and perform better not having to deal with the headache of going
on the road? We seek to answer the questions using scoring totals and shooting
percentages as indicators of team performance. This will deepen understanding
of how home-court advantage affects home and away teams in the NBA.

Our study is quite different from earlier NBA home-court advantage studies.
By using the neutral site games of 2020 we will get to compare home and away
performance to a control. Typically, studies just compare home vs away
performance. These studies do not separate the effects of home-court advantage
into the specific effect on the home team and the specific effect on the away
team. They show that home teams outperform away teams, but not if this is a
result of home teams overperforming or away teams underperforming because of
home-court advantage. Some of these studies are reviewed in greater detail in
Section\textasciitilde\ref{sec:litrev}.

We will compare home team performance in 2020 at a neutral site with
no fans vs.~2017-19 playoffs with fans. Likewise, away team performance in 2020
at a neutral site with no fans vs.~2017-19 playoffs with fans. By comparing home
teams in 2020 to home teams in 2017-19 and away teams in 2020 to away teams in
2017-19, we add a new perspective to the field of research. This will allow for
a more accurate understanding of the effects of home-court advantage on home and
away teams in the NBA. We will not only see that home-court advantage helps home
teams outperform away teams, but also separate the
effects of home-court advantage on home teams and away teams performance individually.

Nine hypotheses were tested to understand the differences in 2020 vs.~earlier
years. First, whether or not the difference between home win percentage in 2020
and 2017-19 is zero. This difference is found to be statistically
significant from zero. Then we assess for differences in home
scoring in 2020 vs 2017-2019. Similarly, we can do the same test, but for
differences in away scoring in 2020 vs 2017-2019. Also, differences in team
shooting (for-two pointers, three-pointers, and free throws) from 2020 vs 2017-2019
for both home and away teams. The results from these tests bring a new perspective
to understanding of how home-court advantage impacts games by altering the performance
of the home and away teams.

\hypertarget{sec:litrev}{%
\section{Literature Review}\label{sec:litrev}}

There is a voluminous literature on the effects of home-court advantage.
Many NBA home-court advantage studies analyze the effects by studying shooting
percentages. \citet{Kotecki} reported significant home-court advantage using performance-based
statistics, specifically field goal percentage, free throw percentage, and
points by comparing home performance vs.~away performance in games.
All of which he showed significantly indicates that
home-court helps teams play better. \citet{Cao} studied the effects
of pressure on performance in the NBA. Using throws as their measure of
interest, they tested whether home
fans could distract and put pressure on opposing players to make free
throws. However, they found insignificant
evidence that home status has a substantial impact on missing from the free throw
line. \citet{Harris} used two point shots, three-point shots and free throws as measures
of interest to study home-court advantage. Two point shots were found to be
the strongest predictor of home-court advantage. They suggested
that home teams should try to shoot more two point shots and force
their opponent to take more two point shot attempts. This strategy will maximize
the benefits of home-court advantage and give them the best chance to win.

Some studies focus less on shooting and more on scoring differences and other metrics.
For example, \citet{Greer} focused on the influence of spectator booing on
home-court advantage in basketball. The three methods of performance used
in this study were scoring, violations, and turnovers. This study was conducted
using the men's basketball programs at two large universities. The study finds that
social support, like booing, is an important contributor to home-court advantage.
Greer explains, whether the influence is greater on visiting team performance or
referee calls is less clear. However, the data does seem to lean slightly in
favor of affecting visiting team performance. \citet{Harville} studied the
effect of home-court advantage using the 1991-1992 college basketball season.
Unlike the NBA, it is not uncommon to have a few games played at neutral sites
during the college basketball season. This allowed them to construct two samples,
one of home teams and one of neutral teams. They formulate their study in a
regression predicting the expected difference in score for home teams.
They set up their study to find if the home teams won games by more points when
they had home-court advantage vs.~when playing at neutral court. This study
concluded with evidence supporting home-court advantage.

There are also surveys on the factors contributing to home-court advantage.
\citet{Carron1992} gave four main game location factors for home and away
teams, namely, the crowd factor, which is the impact of fans cheering; learning
factors, which is an advantage from home teams from playing at
a familiar venue; travel factors, the idea that away teams may face
fatigue and jet lag from traveling; and, rule factors, which says that home teams
may benefit from some advantages in rules and officiating. They acknowledge that
these factors would all be removed if games were played at a neutral site even
if one team was designated as ``home team''. This study was reviewed a decade later
by \citet{Carron2005}. The 2005 review goes over the new findings from studies
about the significance of these four game location factors. Since 1992 they found
that results on these four factors are mixed. However, there is some evidence
supporting crowd and travel factors impact games in the NBA. There is less
evidence suggesting learning and rule factors impact the NBA. One interesting
finding cited by \citet{Carron2005} is that in the absence of crowds result
in overall performance increases.

\hypertarget{sec:data}{%
\section{Data}\label{sec:data}}

Data were collected from the official NBA website. The main variables of interest
are whether or not the home team won, scoring totals for home and away teams, and
shooting percentages for home and away teams on two-pointers, three-pointers, and
free throws. These variables were very popular and frequently used in the related
literature discussed earlier. The data was collected on a game by game basis, this gave us two observations
for each variable per game played, one observation for each team(home and away).
There were 83 games played in the 2020 playoffs, giving 83 observations
for each variable in 2020 for both the home and away teams (166 observations total).
Likewise, there were 243 games played over 2017-2019, giving 243 observations of each variable
for both home and away teams over 2017-2020 (486 observations total).
While many other measures could be used for measuring
the outcome of the game and team performance, scoring seemed to be the most
important. The winner of a game is determined by who scores more points. There can only be
one winner and one loser making the outcome a binary variable, with
one indicating a win and zero indicating a loss.

\begin{figure}
  \centering
  \includegraphics[width=0.9\linewidth]{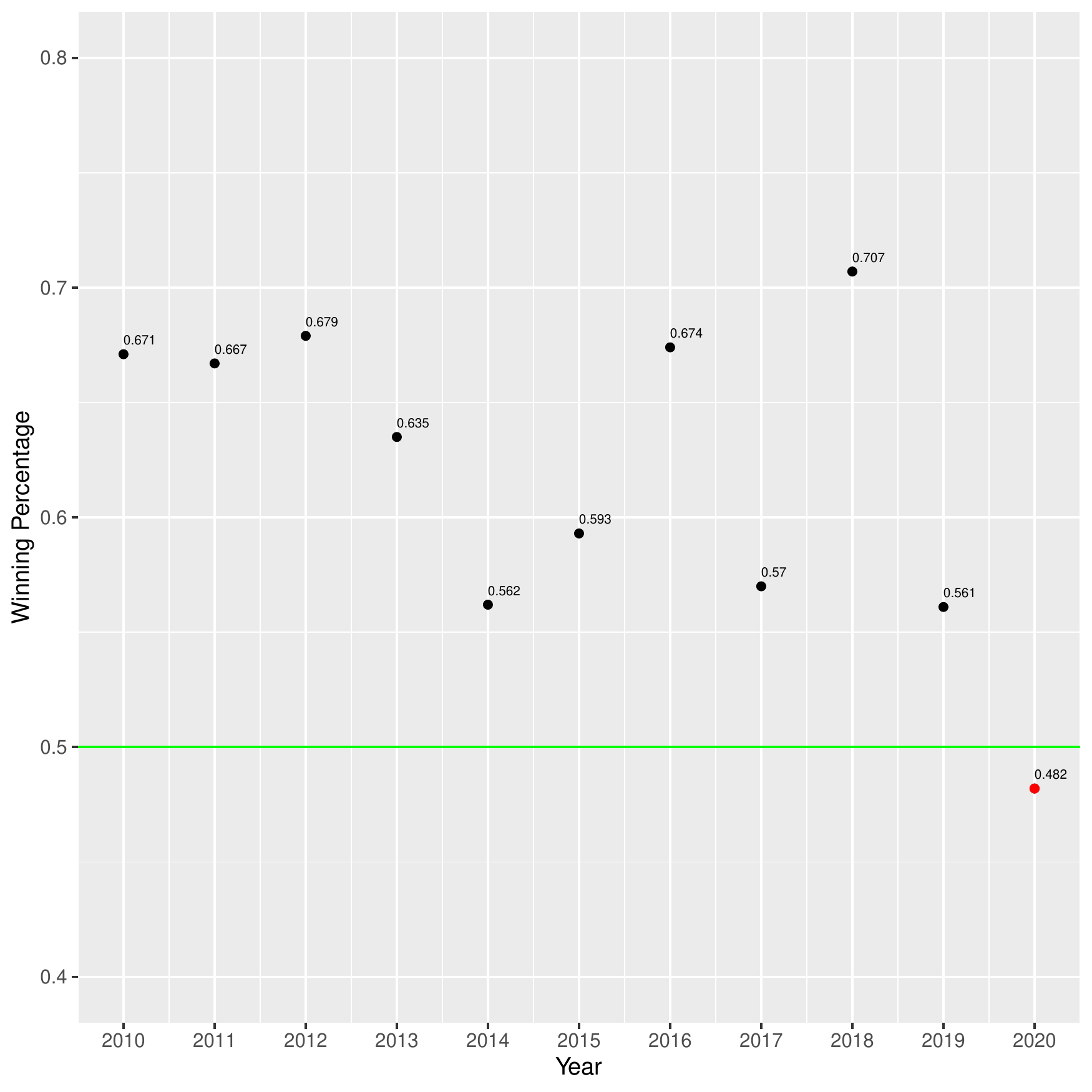}
  \caption{Winning percentage of NBA home teams in the playoffs since
    2010, the green line denotes .500.}
  \label{fig:Fig1}
\end{figure}

Home-court advantage is the basic idea that the home team is more likely to win.
So laying a foundation of typical home-court advantage is crucial. Before
focusing on the 2017 to 2020 playoffs we can take a quick look at home team win
percentages since 2010. Notice in Figure \ref{fig:Fig1}, the 10 years before 2020, the home team
winning percentage ranged from around 0.56 to 0.7 and never dipped below 0.5. The
2020 bubble broke this historic pattern dipping down below 0.5. Foreshadowing,
the confirmation of the expectation that the effect of home-court advantage was removed
in the 2020 playoffs.

\begin{figure}
  \centering
  \includegraphics[width=0.9\linewidth]{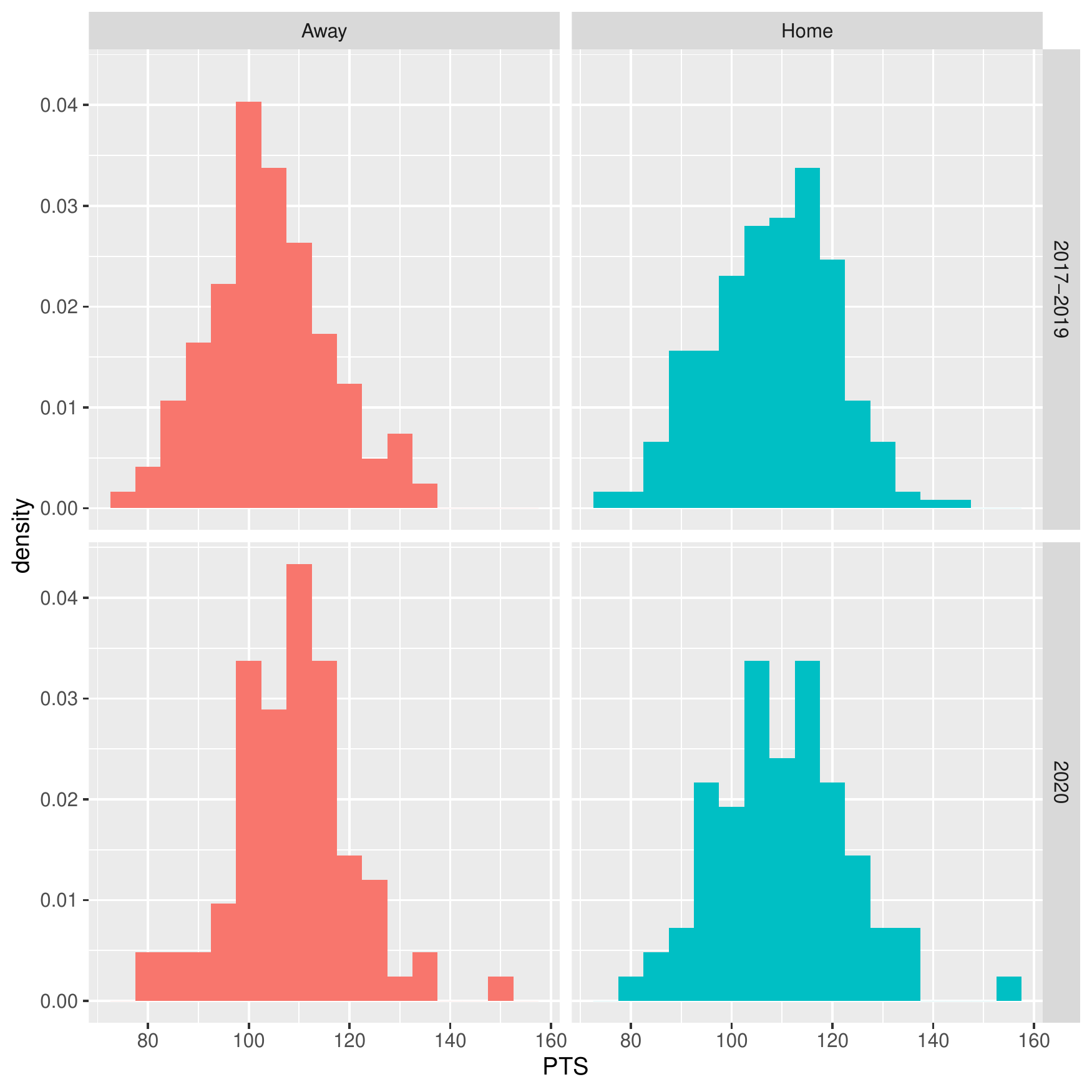}
  \caption{Histograms of home (blue) and away (red) scoring
    for 2020 (bottom) and 2017-2019 (top).}
  \label{fig:Fig2}
\end{figure}

Moving on to the main focus of the study, comparing 2020 to 2017-2019.
Figure \ref{fig:Fig2} shows the histograms of the home (green) and
away scoring (red) for 2020 vs.~2017-2019. All histograms are
fairly bell shaped, which is important for statistical tests
designed for normally distribute data. There appears to be little
difference between the 2020 and 2017-2019 for home scoring, but for
away scoring, a noticeable shift to the right in 2020 is observed
compared with that in 2017-2019.

\begin{figure}
  \centering
  \includegraphics[width=0.9\linewidth]{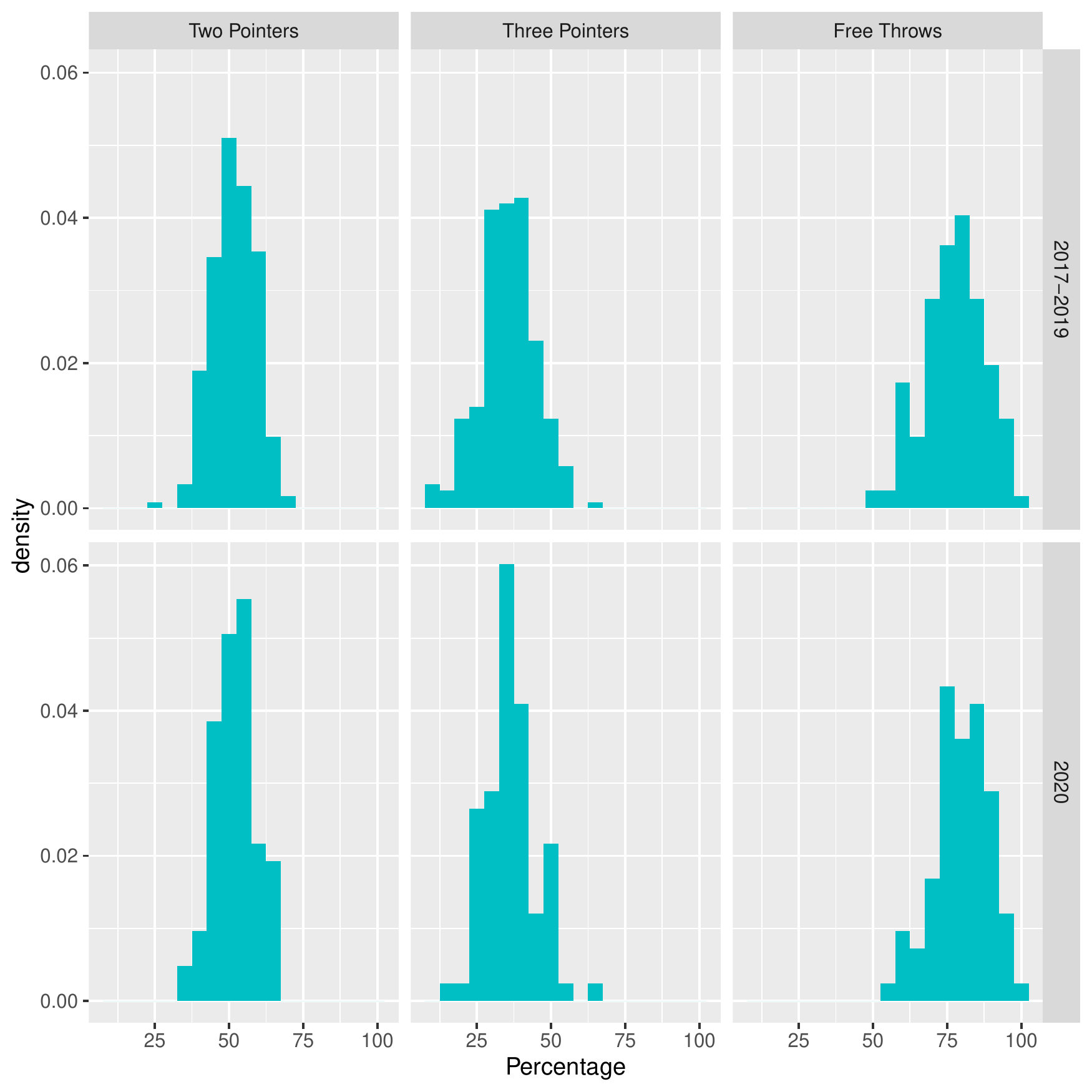}
  \caption{Histograms of home shooting percentages for two-
    pointers, three-pointers and free throws for 2020 (top) vs.~2017-19
    (bottom).}
  \label{fig:Fig3}
\end{figure}

\begin{figure}
  \centering
  \includegraphics[width=0.9\linewidth]{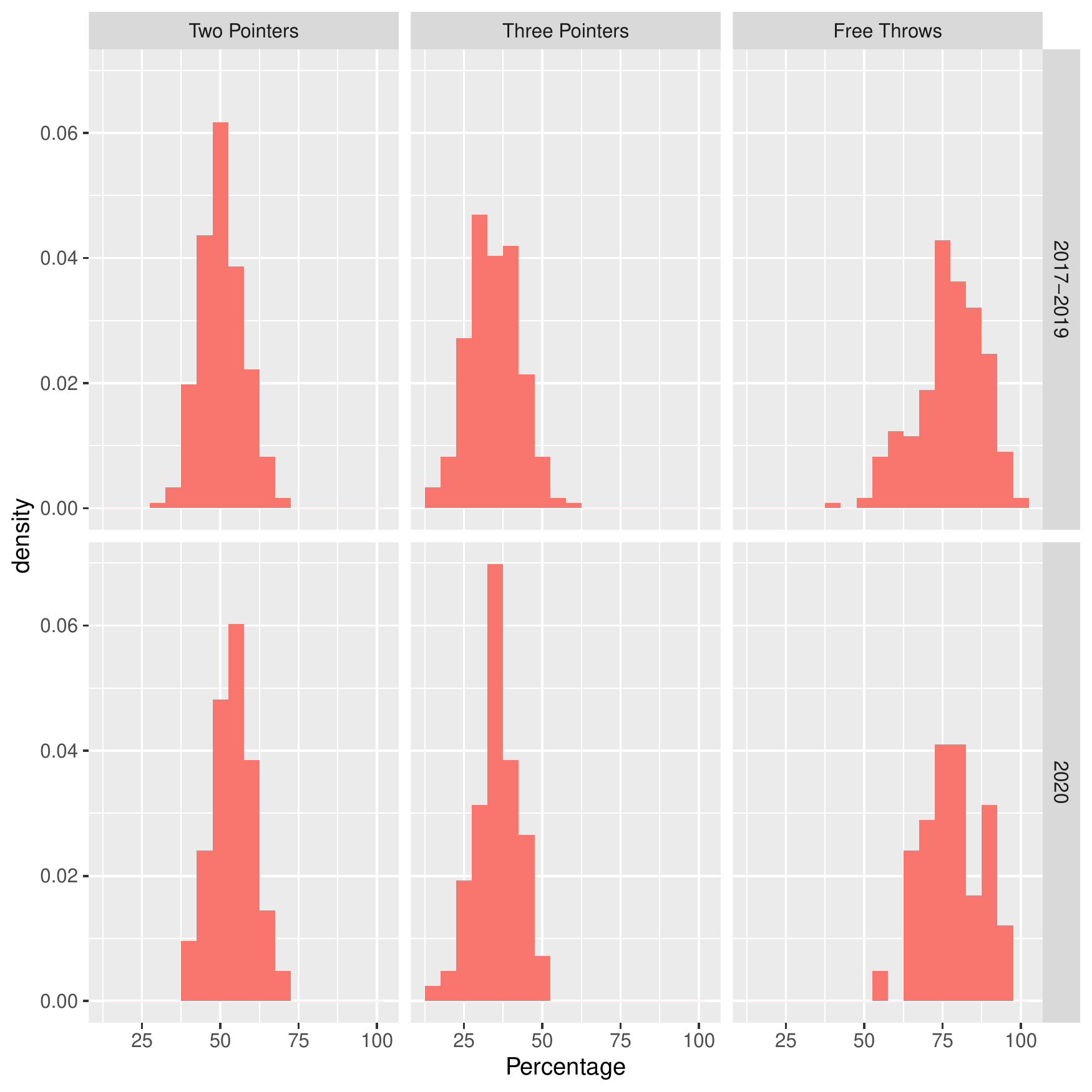}
  \caption{Histograms of away shooting percentages for two-pointers,
    three-pointers and free throws for 2020 (top) vs.~2017-19
    (bottom).}
  \label{fig:Fig4}  
\end{figure}

Our second target of inference is shooting percentage for home and away teams.
Figure \ref{fig:Fig3} shows home shooting for two-pointers, three-pointers and
free throws for 2020 (top) vs.~2017-19 (bottom). The histograms appear to be fairly
similarly distributed between 2020 and 2017-19. Likewise, Figure \ref{fig:Fig4},
shows the same percentages except for away teams. It appears that the
two-point shooting percentage for away teams has a small shift to the
right in 2020 relative to that in 2017-2019.

\hypertarget{sec:methods}{%
\section{Methods}\label{sec:methods}}

The 2020 bubble provides a new and exciting opportunity to study home-court
advantage for the NBA. Unlike college basketball, aside from a few
exhibition/preseason games, the NBA always has a home and away team. So, for
the first time in NBA history the bubble allows NBA home and away performance
to be compared vs a control/neutral field. The NBA bubble, as a neutral court,
removed all 4 possible game location factors impacting home-court advantage
hypothesized by \citet{Carron1992}. The NBA bubble featured 8 seeding games then a
standard playoff format. The focus of this study was the play during the playoff
games since it followed the standard playoff format and can easily be compared
back to other playoffs. For this study, the 2020 playoffs were compared against
the three previous playoffs collectively. To control
for the changing play style of the NBA, we limit the study to 2020 vs 2017-2019
for the faster pace play and more common use of the three-point shot
in modern basketball. If we used data from say 10 years ago, or earlier, observed
differences may not be from effects of the NBA bubble, but rather from the effects
of drastic changes in the style of play between the seasons. However, basketball
evolves slow enough that we can reasonably assume 2017-2019 are at least very
close in pace and playing style to 2020.

Comparisons between 2020 and 2017-19 home and away teams were made on home team
winning percentage, total team scoring and two-point, three-point and free throw
shooting. Comparing the differences in these metrics for home and away teams in
2020 vs previous years will provide valuable insights to the understanding of
home-court advantage. We can see how going on the road may negatively impact
away performance and how playing at home may positively impact home performance.
If there are differences in scoring for home or away teams the differences can
be used to show how home-court advantage affects overall performance of home
and away teams. While testing for differences in shooting will provide added
context for how home-court advantage specifically affects performance.
Shooting percentages are not the only possible metrics affected by home-court
advantage, but they are the most obvious and certainly an important one.

We formulate the following nine specific research questions to test
the effects of the COVID bubble on the 2020 NBA playoffs:

\begin{enumerate}
\def\labelenumi{\arabic{enumi}.}
\item
  Is the home team winning percentage in 2020 different than that it was in 2017-2019?
\item
  Is the average home team scoring different in 2020 than it was over 2017-2019?
\item
  Is the average away team scoring different in 2020 than it was over 2017-2019?
\item
  Are home teams making two-pointers at the same rate in 2020 as 2017-2019?
\item
  Are home teams making three-pointers at the same rate in 2020 as 2017-2019?
\item
  Are home teams making freethrows at the same rate in 2020 as 2017-2019?
\item
  Are away teams making two-pointers at the same rate in 2020 as 2017-2019?
\item
  Are away teams making three-pointers at the same rate in 2020 as 2017-2019?
\item
  Are away teams making free throws at the same rate in 2020 as 2017-2019?
\end{enumerate}

All nine questions can be approached by a standard two-sample
comparison with the \(z\)-test. The \(z\)-test statistic follows a
standard normal distribution, which is a good approximation based on
the central limit theorem given the sample size in this application.

We also conducted nonparametric tests that are distribution free to
confirm the results from the \(z\)-test. For question\textasciitilde1, we used
Fisher's exact test for a contigency table which summarizes the wins
and losses of the home team in the 83 games in 2020 and the 243 games
in 2017-2019. For all other eight questions, the data are the scores
or shooting percentages from the 84 games in 2020 and the 243 games in
2017-2019. We used Wilcoxan's rank-sum test.

All three tests, namely the \(z\)-test, Fisher's exact test, and
Wilcoxan's rank-sum test, were performed using R \citep{R}.

\hypertarget{sec:results}{%
\section{Results}\label{sec:results}}

\begin{table}
  \caption{The results from the 9 tests.}
  \label{tab:table}
\centering
\begin{tabular}[t]{lcccc}
  \toprule
  & 2020 & 2017-19 & \multicolumn{2}{c}{P-value}\\
  \cmidrule(lr){4-5}
  &          &                & Z-test & Non-parametric\\
\midrule
Home Win & 0.482 & 0.613 & 0.0497 & 0.0400\\
Home Scoring & 1.101 & 1.081 & 0.2321 & 0.2985\\
Away Scoring & 1.091 & 1.040 & 0.0008 & 0.0004\\
Home 2P & 0.523 & 0.515 & 0.4335 & 0.5719\\
Home 3P & 0.363 & 0.357 & 0.5733 & 0.8852\\
Home FT & 0.793 & 0.774 & 0.0692 & 0.0496\\
Away 2P & 0.536 & 0.504 & 0.0003 & 0.0003\\
Away 3P & 0.357 & 0.346 & 0.3256 & 0.3081\\
Away FT & 0.783 & 0.777 & 0.6601 & 0.8370\\
\bottomrule
\end{tabular}
\end{table}

Starting from the top of the
Table \ref{tab:table} summarizes p-values of the nine hypotheses for both
\(z\)-tests and Wilcoxon tests. The p-values are all fairly similar for both
tests giving strong confidence in conclusions drawn. Also reported are
the point estimates of the two samples in each comparison.

First, we see a statistically significant change
in home win percentage in 2020 from 2017-19, with p-value of 0.0497 for the
\(z\)-test and 0.0400 for Fisher's exact test. The 95\% confidence
interval (CI) of \((-0.255, -0.008)\)
confirms our belief that home-court advantage was lost in the 2020 NBA playoffs. However, after accounting for
multiple tests using the Bonferroni correction, the p-values for both tests are no longer significant.
So, we may only cautiously say there is evidence that home-court advantage was
not a factor in 2020.

Home team performance did not seem to be negatively impacted by losing home-court
advantage like expected. Home scoring, two-point and three-point shooting
shooting all show no significant difference, on average, between 2020 vs.
2017-19 based on p-values from both tests. However, the Wilcoxon test and \(z\)-test have conflicting results
for free throws. The \(z\)-test p-value of 0.0692 indicates no significant difference, while
the Wilcoxon test p-value of 0.0469 indicates a difference at the 5\% significance level. Since,
the p-value of Wilcoxon test is so close to significance level and neither p-value is significant
after a Bonferonni correction for multiple tests this difference is likely not very meaningful.
There appears to be no strong evidence suggesting home teams played at
a lower level in 2020 than they did in previous years when they had home-court advantage.

Away teams saw more of an impact than home teams. For starters, there is a
significant increase in mean points per game, indicated by p-value of 0.0008 for \(z\)-test and
0.0004 for Wilcoxon. It is important to note both p-values also remain significant after
a Bonferonni correction giving strong indication of significance. The average difference in points was estimated to
be about 5 points, with 95\% CI \((2.083, 7.988)\). Likewise, the away team two-point
shooting efficiency increased significantly based on p-value of 0.0003
for both the \(z\)-test and Wilcoxon test. Again, both p-values remain significant after Bonferroni correction.
The average difference was estimated to be about 0.03, with 95\% CI \((0.015, 0.050)\). However,
unlike two-point shooting, away teams did not see a statistically significant difference in three-point and free
throw shooting. Overall, away teams have evidence of change in performance in
the bubble. The away teams seemed to perform better than they would under normal
conditions as a visiting team.

\hypertarget{sec:disc}{%
\section{Discussion}\label{sec:disc}}

Generally it seemed that away teams fared better in the 2020 NBA playoff bubble
than previous years on the road. Starting from the dip in home winning percentage
to below 0.482 it is clear that something was different. Although the difference was not
significant after a Bonferroni correction it is still informative to consider and
understand that home teams seemed to struggle to win compared to normal conditions.
Compared to \citet{Kotecki} who finds home teams consistently have a significantly
better record than away teams boasting about a 60.5\% win percentage in his sample,
the 48.2\% home winning percentage of 2020 home teams is quite a shift.
In this study, home teams did not benefit from the usual advantages provided by
being the home team.

Away team average
scoring did increase by a statistically significant amount. This goes hand in
hand with our intuition and conclusion about the home winning percentage
decreasing. If away teams are scoring significantly more and home teams are not,
then we expect to see away teams winning a larger amount of games. This may give
more reason to believe the conclusion that there was a significant decrease in
home winning percentage in 2020, despite failing to be significant
after the Bonferroni
correction. Only away team scoring being significantly impacted by playing on
a neutral court indicates that home-court advantage stems mainly from adverse
effects on the visiting team.

An interesting finding is all shooting and scoring numbers for both home and
away teams did make at least small increases. Although these increases
were not
all significant these increases are exactly what is reported in \citet{Carron2005}
when they explain how evidence suggests that teams perform better with the absence
of fans. This is important because it coincides with our conclusion that home-court
advantage mostly plays into games by negatively impacting away teams. If fans cause
overall performance to drop, then home court advantage must come from a bigger drop
in away performance than the drop in home performance. This is why away teams were
able to close the gap with home teams with home-court advantage removed.

Separating the effects of the home-court advantage into home effects and away
effects allowed for some interesting new insights. Previously, we knew that on
average home teams outperformed away teams. It was less clear whether it was
from positive effects on the home team or negative effects on the road team or
perhaps a bit of both. The biggest takeaway from this study is the main source
of home-court-advantage is the negative effects playing on the road away teams
face. In 2020 there wasn't any evidence of regression for home team performance,
based on the performance measures used, despite being stripped of home-court
advantage. Yet, home teams lost about 12\% more of games in the 2020 playoffs
than the typical average. This was because of the improvement of away teams.
No longer having to face the struggle of traveling, pressure from opposing fans,
or playing on an unfamiliar court, teams saw an improvement in their play and an
increase in winning. The improvement of away teams confirms a proposition from
\citet{Greer} that the positive social impact of crowds benefiting
home teams may be a result of inhibiting away teams.

At least some of that improvement from away teams came from significantly higher
two-point efficiency. This corresponds with the conclusion from \citet{Harris},
where they found home teams are best suited to capitalize on advantages from two-point
shots. Normally, by shooting more two-pointers themselves and forcing away teams to
shoot more two-pointers the home team benefits from increasing effects of
home-court advantage. However, with away teams significantly improving two-point
shooting in the bubble this strategy was no longer viable and home-court
advantage disappeared.

Future studies may want to use the 2020 NBA bubble and compare vs previous years
using other performance measures. For example, turnovers, steals, assist, rebounds,
and many more game statistics. There are plenty of other possibilities besides
just shooting efficiency to pick through looking for more possible sources of
added points for away teams. This will further help explain what is lost in the
performance of away teams when they travel to opposing arenas. This study is only
the beginning of possibilities for studies using the 2020 NBA bubble as a case study
for home-court advantage. Although the study is limited by a one time
sample, it seems unlikely that these conditions will ever be
repeated. It may not be possible to have a follow-up study using the
same measures with a different sample. Otherwise, that type of study
could help strengthen the conclusion in this paper.

\bibliographystyle{chicago}
\bibliography{citations.bib}

\end{document}